\documentclass[letterpaper]{article} 
\usepackage{aaai2026}  
\usepackage{times}  
\usepackage{helvet}  
\usepackage{courier}  
\usepackage[hyphens]{url}  
\usepackage{graphicx} 
\usepackage{natbib}  
\usepackage{caption} 
\usepackage{algorithm}
\usepackage{algorithmic}
\usepackage{multirow}
\usepackage{amsmath}
\usepackage{booktabs}
\usepackage{amsfonts}
\usepackage{subcaption}
\usepackage{xcolor}

\usepackage{newfloat}
\usepackage{listings}
\DeclareCaptionStyle{ruled}{labelfont=normalfont,labelsep=colon,strut=off} 
\floatstyle{ruled}
\newfloat{listing}{tb}{lst}{}
\floatname{listing}{Listing}
\nocopyright

\title{RAP: Real-time Audio-driven Portrait Animation with Video Diffusion Transformer}
\author{
    Fangyu Du\textsuperscript{\rm 1,2}\equalcontrib,
    Taiqing Li\textsuperscript{\rm 1,3}\equalcontrib,
    Qian Qiao\textsuperscript{\rm 1},
    Tan Yu\textsuperscript{\rm 1},
    Ziwei Zhang\textsuperscript{\rm 1},
    Dingcheng Zhen\textsuperscript{\rm 1},    
    Xu Jia\textsuperscript{\rm 3},
    Yang Yang\textsuperscript{\rm 2}, 
    Shunshun Yin\textsuperscript{\rm 1}$^\dagger$,
    Siyuan Liu\textsuperscript{\rm 1}\thanks{Corresponding author, Siyuan Liu is the Project Leader.}
}
\affiliations{
    \textsuperscript{\rm 1}Soul AILab, \textsuperscript{\rm 2}Xi'an Jiaotong University, \textsuperscript{\rm 3}Dalian University of Technology\\

    \texttt{zjdfy2019@stu.xjtu.edu.cn,\small\{qiaoqian, yutan, liusiyuan\}} \\
  \texttt{\small@soulapp.cn} \\
  Project: \url{https://qianqiaoai.github.io/RAP/} 
}

\usepackage{bibentry}

\begin{document}

\maketitle

\begin{abstract}

Audio-driven portrait animation aims to synthesize realistic and natural talking head videos from an input audio signal and a single reference image. While existing methods achieve high-quality results by leveraging high-dimensional intermediate representations and explicitly modeling motion dynamics, their computational complexity renders them unsuitable for real-time deployment. Real-time inference imposes stringent latency and memory constraints, often necessitating the use of highly compressed latent representations. However, operating in such compact spaces hinders the preservation of fine-grained spatiotemporal details, thereby complicating audio-visual synchronization and increasing susceptibility to temporal error accumulation over long sequences. To reconcile this trade-off, we propose \textbf{RAP} (\textbf{R}eal-time \textbf{A}udio-driven \textbf{P}ortrait animation), a unified framework for generating high-quality talking portraits under real-time constraints. Specifically, RAP introduces a hybrid attention mechanism for fine-grained audio control, and a static-dynamic training-inference paradigm that avoids explicit motion supervision. Through these techniques, RAP achieves precise audio-driven control, mitigates long-term temporal drift, and maintains high visual fidelity. Extensive experiments demonstrate that RAP achieves state-of-the-art performance while operating under real-time constraints.

\end{abstract}
\section{Introduction}

Recent advances in latent diffusion models have significantly improved the controllability and quality of portrait animation. Among various control modalities, audio offers continuous and fine-grained temporal cues that naturally align with facial dynamics, making it particularly suitable for driving talking portrait generation. Leveraging these properties, recent audio-driven methods~\cite{xu2024hallo,chen2025echomimic} have achieved impressive progress in synthesizing realistic facial expressions and accurate lip movements, greatly enhancing the expressiveness of animated portraits. However, as portrait animation becomes increasingly integrated into interactive scenarios such as virtual communication, digital avatars, and live-streaming, achieving low-latency, high-fidelity generation is critical for delivering smooth and responsive user experiences.

To further improve generation quality and temporal consistency, recent works such as the Hallo~\cite{xu2024hallo,cui2024hallo2,cui2025hallo3} and EchoMimic~\cite{chen2025echomimic,meng2025echomimicv2} series introduce sophisticated design strategies. The Hallo series employs dynamic masking and fixed noise injection to guide the model's attention toward facial regions, effectively decoupling local details from global semantics. The EchoMimic series adopts a multi-stage training paradigm that first captures coarse-grained motion and then refines high-frequency details, while incorporating pose control mechanisms to reduce motion drift.

\begin{figure}[t]
    \includegraphics[width=\linewidth]{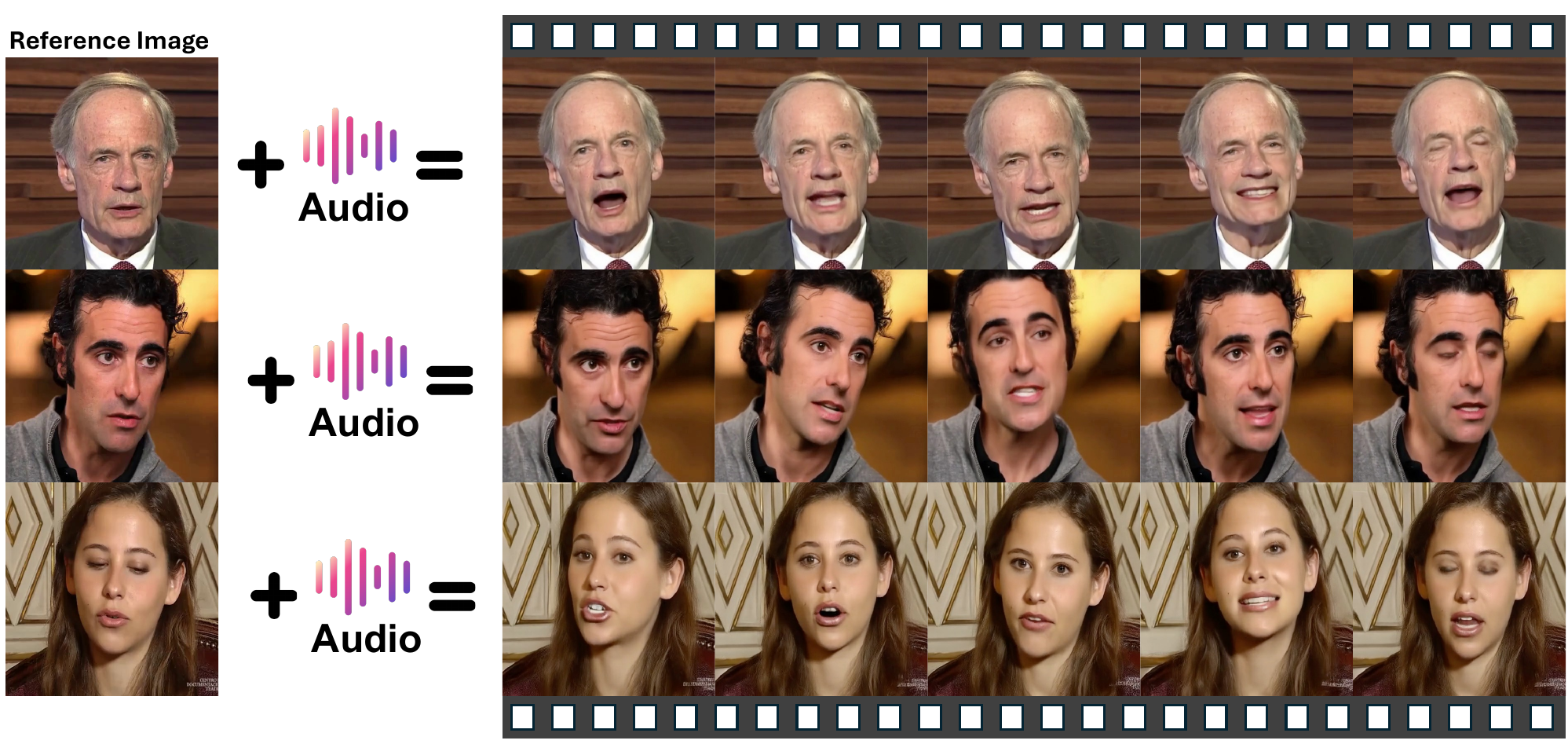}
    \caption{Illustration of the proposed portrait animation framework RAP. Given a reference image and an audio clip, the model generates a natural and vivid talking portrait.}
    \label{fig:headimage-V1}
\end{figure}

Although effective, existing methods often rely on high-dimensional representations or fine-grained visual storage to maintain coherence and identity consistency. However, this leads to high computational and memory costs, limiting their applicability in real-time settings.
Real-time, high-quality audio-driven portrait animation remains challenging due to two main issues: (1) Fine-grained control under high compression. Techniques like LTX-VAE~\cite{hacohen2024ltx} speed up inference by reducing token length, but increase information density per token, imposing stronger requirements on the diffusion model's ability for fine-grained control. (2) Error accumulation in long sequences. Small prediction errors gradually build up over time, causing motion discontinuities, identity drift, and image distortion in longer sequences.

To address the above challenges, we propose RAP, a unified real-time portrait animation framework. RAP aims to simultaneously achieve real-time, high-quality inference performance while effectively mitigating error accumulation issues in long-term generation. Our method is built upon high compression ratio spatiotemporal latent representations to meet strict real-time requirements. To mitigate the difficulty of fine-grained control imposed by highly compressed latent spaces, we innovatively design a hybrid attention mechanism. This mechanism cleverly combines attention to global video coherence with precise control of audio features on fine-grained temporal dimensions over key local video regions (e.g., mouth, eyes). This significantly improves the quality of generating local details under high compression latent space, particularly enhancing the precision of audio-video synchronization.

Furthermore, to effectively solve error accumulation and identity drift issues in long-term generation, RAP proposes a training and inference strategy without explicit motion frame storage. Its core is the innovative static-dynamic hybrid training paradigm. Under this paradigm, we distinguish between static latents and dynamic latents in the VAE latent space. To ensure consistent training and inference, the model is trained to initiate generation both from static latents (for the first clip) and from dynamic latents (for subsequent clips). Rather than applying hard conditioning on preceding outputs, which may lead to the accumulation of errors over time, we adopt a soft guidance mechanism that reuses the denoising process of prior latent features. Through this method, RAP can achieve nearly infinite-length real-time generation while maintaining ID and detail features.

In summary, the main contributions of this paper can be summarized as follows:
\begin{itemize}
\item We propose RAP, a novel audio-driven real-time portrait animation generation framework that can generate high-quality, realistic portrait animations while ensuring efficient inference.

\item To meet the precision demands of highly compressed latent spaces, we propose a hybrid attention mechanism. It effectively fuses global video context with fine-grained audio cues, enhancing local detail generation and audio-video synchronization.

\item To address error accumulation and identity drift in long video generation, we propose a static-dynamic hybrid paradigm with soft latent guidance to support seamless, long-term video generation without explicit motion conditioning.

\item We conduct extensive experiments and in-depth analyses to comprehensively validate the effectiveness of the proposed RAP framework. Results demonstrate that RAP achieves high-quality portrait animation and strong temporal coherence under real-time constraints.

\item To promote further research in the real-time portrait animation generation field, we will open-source our data cleaning and processing pipeline, as well as complete model training and inference code. 
\end{itemize}

\section{Related Work}

\begin{figure*}[t]
     \centering
    \includegraphics[width=\linewidth]{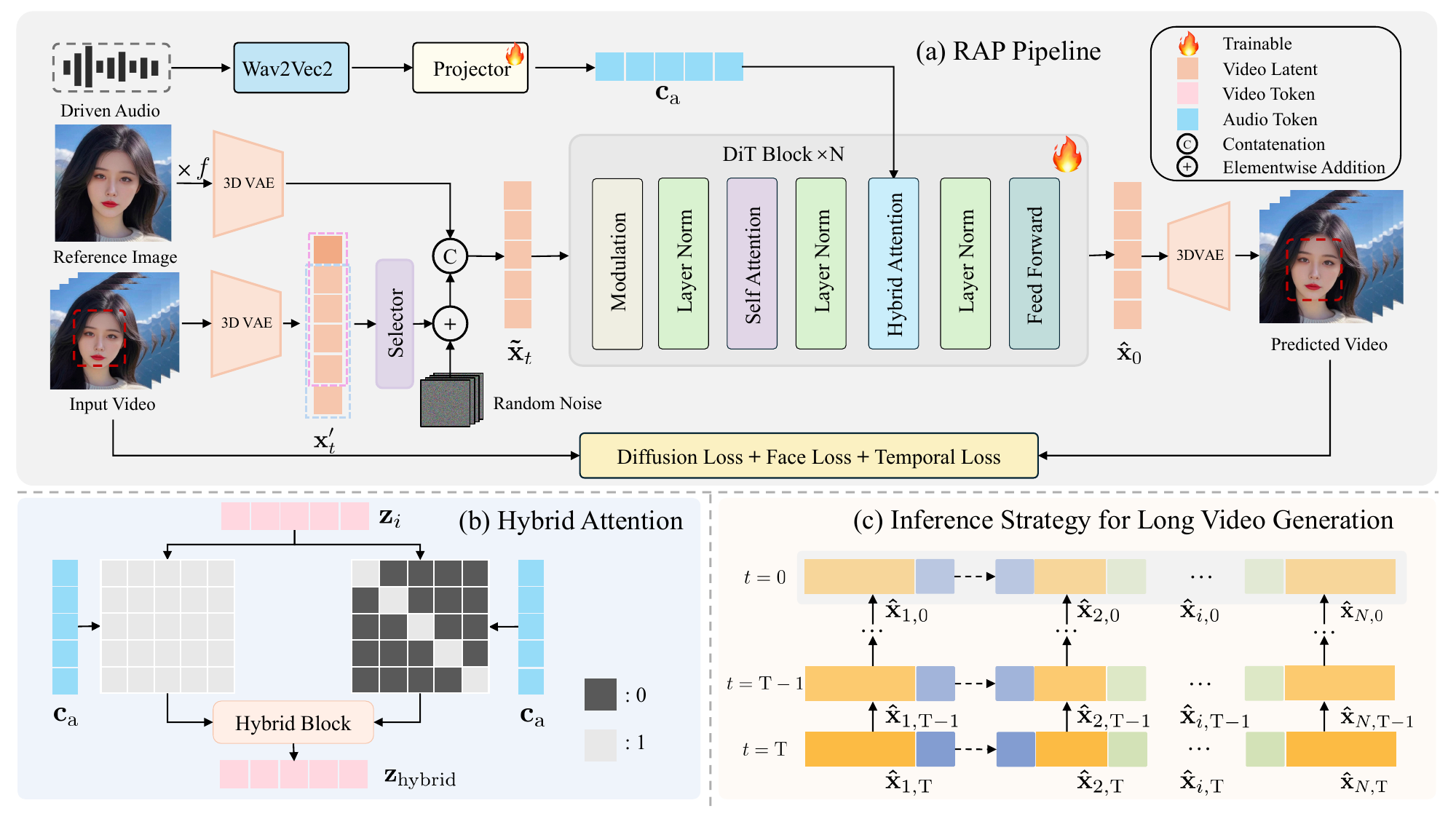}
    \caption{Overview of the proposed RAP framework. (a) Overview of the RAP pipeline, where audio and image inputs are encoded into compressed tokens, followed by DiT-based denoising to generate high-quality portrait videos. (b) The hybrid attention block conducts cross-attention at both short-term and long-term temporal scales, and fuses the results to capture multi-scale audio-motion dependencies. (c) A step-wise inference strategy that progressively guides video generation in the latent space by inheriting information across timesteps. }
    \label{fig:RAP-V1}
\end{figure*}

\subsubsection{Diffusion Models for Video Generation.}

Diffusion models~\cite{ho2020denoising,song2020denoising} have shown strong potential in video generation by modeling complex spatial-temporal dynamics. Early approaches mainly use UNet-based architectures~\cite{blattmann2023stable,bar2024lumiere,singer2022make,wang2023modelscope,wang2024recipe,wu2023tune,chai2023stablevideo,ceylan2023pix2video,guo2023animatediff}, where temporal modules are added on top of spatial UNets. However, this separation limits interaction between spatial and temporal features. As a result, these models tend to produce temporal artifacts like frame-wise inconsistencies or flickering pixels, especially in dynamic or textured regions of the video.

To address these limitations, recent advances have shifted toward DiT-based architectures~\cite{yang2024cogvideox,kong2024hunyuanvideo,wan2025wan}, which treat video frames as spatio-temporal tokens and model them using global self-attention across both dimensions. This unified token-based formulation enables tighter integration of spatial and temporal information, leading to more coherent motion and better generalization to complex video generation tasks. DiT frameworks naturally support cross-modal conditioning—such as text, audio, or pose—by embedding additional control tokens into the same attention space, resulting in superior controllability and alignment. These architectural improvements have contributed to substantial gains in generation quality, positioning DiT-based diffusion models as the current state-of-the-art for high-quality video synthesis.

\subsubsection{Audio-driven Portrait Animation.}
Portrait animation generation aims to produce realistic and temporally coherent facial motions that preserve identity and expression. Audio has become a widely used driving signal due to its rich temporal cues and strong correlation with speech-related dynamics. Early methods~\cite{cheng2022videoretalking,wang2023seeing,gan2023efficient} often rely on GANs or explicit motion models. For example, Wav2Lip~\cite{prajwal2020lip} ensures accurate lip sync via a dedicated discriminator, while SadTalker~\cite{zhang2023sadtalker} maps audio to 3DMM parameters for rendering. However, these methods often struggle with expressiveness and temporal consistency, especially in longer sequences.

Recent diffusion-based approaches significantly improve generation quality and coherence. UNet-based models~\cite{xu2024hallo,cui2024hallo2,tian2024emo,wei2024aniportrait,chen2025echomimic} learn audio-to-video mappings in a data-driven manner, enabling more expressive facial motion. For long-form generation, MEMO~\cite{zheng2024memo} and LOOPY~\cite{jiang2025loopytamingaudiodrivenportrait} model inter-clip dependencies to maintain motion continuity. DiT-based methods~\cite{cui2025hallo3,wang2025fantasytalking,lin2025omnihuman} further enhance spatial-temporal coherence and audio-visual alignment through global attention and token-level control. However, their high computational cost remains a challenge for real-time use.

To address these challenges, we propose RAP (Real-time Audio-driven Portrait Animation). It is a high-performance framework tailored for real-time generation under high compression. RAP employs a hybrid attention mechanism and a static-dynamic training-inference strategy that improve spatiotemporal consistency while keeping latency low. 

\section{Methodology}
This section systematically presents the methods used in our study. It is organized into four parts: the preliminaries, the RAP framework, a Hybrid Attention mechanism to enhance audio-visual alignment, and a unified training and inference strategy to ensure stable and coherent results.

\subsection {Preliminary}

\paragraph{Diffusion Transformer (DiT).}
DiT~\cite{peebles2023scalable} is a generative model that combines Transformer architectures with diffusion processes, aiming to overcome the structural limitations of traditional U-Net-based latent diffusion models (LDMs)~\cite{rombach2022high}. By replacing the U-Net~\cite{ronneberger2015u} with a Transformer, DiT enhances modeling capacity and scalability across both spatial and temporal dimensions.

In video generation, DiT is often combined with a causal 3D VAE~\cite{kingma2013auto,zheng2024open} for spatio-temporal compression. Conditional inputs (e.g., text) are incorporated via adaptive normalization or cross-attention to enable controllable generation.
The core objective of DiT is to learn a vector field by minimizing the mean squared error (MSE) between the predicted and ground-truth velocity fields. The training objective is:
\begin{equation}
\label{flow_matching1}
\mathcal{L}_{\text{FM}}(\theta) = \mathbb{E}_{\mathbf{t},\, p_{\mathbf{t}}(\mathbf{x})} \left\| \mathbf{v}_t - \mathbf{u}_t \right\|^2,
\end{equation}
where $\boldsymbol{\theta}$ denotes the model parameters, $\mathbf{x}_t \sim p_t(\mathbf{x})$ is a sample at time step $t$, $\mathbf{v}_t$ is the predicted velocity field, and $\mathbf{u}_t = \frac{d\mathbf{x}_t}{dt}$ is the ground-truth velocity.
Here, $\mathbf{x}_t$ is obtained by linearly interpolating between a real sample $\mathbf{x}_0$ and Gaussian noise $\mathbf{x}_1 \sim \mathcal{N}(\mathbf{0}, \mathbf{I})$, as
\begin{equation}
\mathbf{x}_t = t \cdot \mathbf{x}_1 + (1 - t) \cdot \mathbf{x}_0.
\end{equation}
We adopt the Wan2.1 Text-to-Video (T2V) model~\cite{wan2025wan} with 1.3 billion parameters as our baseline, offering a strong trade-off between performance and computational efficiency in our video generation pipeline.

\paragraph{LTX-VAE.}

LTX-VAE~\cite{hacohen2024ltx} is a 3D video VAE that maintains high performance at high compression ratios.
To reduce the quadratic computation cost of attention, LTX-VAE downsamples the RGB input with a factor of \( (32, 32, 8) \) in space and time, achieving a pixel-to-token ratio of \( 1\!:\!8192 \). This is a \( 32\times \) higher compression than the commonly used \( 8\times8\times4 \) scheme, significantly improving efficiency. To compensate for detail loss from heavy compression, LTX-VAE incorporates the final step of a diffusion model in the decoder, enhancing visual quality. To enable real-time video generation, we adopt LTX-VAE in our system. 
However, although LTX-VAE maintains a comparable static frame performance while drastically reducing time, it still lacks in dynamic frames and temporal continuity, and the increase in the number of frames mapped by its single token increases the difficulty of aligning the details at the frame level in the subsequent network.
\subsection{RAP}

We propose a Real-time Audio-driven Portrait animation model called RAP, which uses a reference image \( \mathbf{I} \) and an audio clip \( \mathbf{A} \) to generate an identity-consistent portrait animation \( \hat{\mathbf{V}} \), as illustrated in Figure~\ref{fig:RAP-V1}. 

Let \( \mathbf{x}_t \) denote the noisy video latent at timestep $t$. To incorporate identity information, the reference image \( \mathbf{I} \) is temporally repeated and then encoded by a variational autoencoder \( \mathcal{E} \), yielding a latent representation \( \mathbf{x}_{\text{ref}} \) that is channel-aligned with \( \mathbf{x}_t \). The two latents are then concatenated along the channel dimension to form a fused representation \( \tilde{\mathbf{x}}_t = \mathrm{Concat}(\mathbf{x}_t, \mathbf{x}_{\text{ref}}) \).

The audio clip A is encoded by a pretrained Wav2Vec2 model~\cite{baevski2020wav2vec}, and then passed through a multi-layer perceptron (MLP) to extract temporally aligned audio features
$\mathbf{c}_\text{a} = \mathrm{MLP}(\mathrm{Wav2Vec2}(\mathbf{A}))$.

Finally, the RAP model \( \mathcal{M} \) takes $\tilde{\mathbf{x}}_t$, \( t \), and \( \mathbf{c}_\text{a} \) as inputs, and predicts the velocity field \( \mathbf{v}_t \), which guides the denoising trajectory and ultimately leads to the generation of the audio-driven portrait video. To optimize the model, we propose a composite Flow Matching loss comprising three terms:


\begin{equation}
\label{eq:flow_matching}
\begin{aligned}
\mathcal{L} = \mathbb{E}_{\mathbf{t},\, p_{\mathbf{t}}(\mathbf{x})} \bigg[
& \left\| \mathbf{v}_t - \mathbf{u}_t \right\|^2
+ \lambda \left\| \mathbf{m} \odot (\mathbf{v}_t - \mathbf{u}_t) \right\|^2 \\
& \hfill + \mu \left\| \Delta \mathbf{v}_t - \Delta \mathbf{u}_t \right\|^2 
\bigg],
\end{aligned}
\end{equation}
where \( \mathbf{v}_t \) and \( \mathbf{u}_t \) denote the predicted and ground-truth velocity fields, \( \mathbf{m} \) is a facial region mask, and \( \odot \) represents element-wise multiplication. The first term (Diffusion Loss) enforces overall motion accuracy, the second term (Face Loss) emphasizes facial motion fidelity, and the third term (Temporal Loss) enforces temporal consistency by minimizing velocity differences across adjacent frames, with
$\Delta \mathbf{v_t} = \mathbf{v_t}{[:,1:]} - \mathbf{v_t}{[:,:-1]},\ \Delta \mathbf{u_t} = \mathbf{u_t}{[:,1:]} - \mathbf{u_t}{[:,:-1]}.$
The weights \( \lambda \) and \( \mu \) control the contribution of the face-focused and temporal regularization terms.

\subsection{Hybrid Attention}
Audio, as a temporally aligned conditional signal, carries rich semantic information. It explicitly governs lip movements at the frame level and implicitly affects global facial expressions and motion intensity. Audio influences these aspects at different temporal scales: lip synchronization requires frame-level alignment, while expressions and motions change more gradually. In high-compression VAEs, each latent spans multiple frames, challenging fine-grained lip alignment alongside overall facial dynamics. To address this, we propose a hybrid attention module that fuses audio and visual latent features at both full-sequence and local-region scales, enabling precise and comprehensive control during generation.

Specifically, we denote the input video tokens to the \( i \)-th DiT block as \( \mathbf{z}_i \in \mathbb{R}^{(F \times H \times W) \times D} \), where \( F \) is the number of latent frames, each frame is divided into \( H \times W \) spatial patches, and \( D \) denotes the feature dimension used in the Transformer. For simplicity, we omit the time step \( t \) in \( \mathbf{z}_i \). We patchify the fused latent \( \tilde{\mathbf{x}}_t \) to obtain the initial video tokens \( \mathbf{z}_0 \in \mathbb{R}^{(F \times H \times W) \times D} \). Meanwhile, the audio features \( \mathbf{c}_\text{a} \in \mathbb{R}^{(F \times r_f \times N) \times D} \) are extracted as described above, where \( r_f \) is the temporal compression ratio of the VAE and \( N \) is the number of audio feature layers.
These two sequences serve as inputs to the \( i \)-th DiT block, where we design two cross-modal fusion mechanisms based on cross-attention~\cite{vaswani2017attention}.

\paragraph{Full-Sequence Fusion.}
The global fusion output \( \mathbf{z}_{\mathrm{full}} \) is obtained by applying a global cross-attention between the entire sequence of \( \mathbf{z}_i \) and \( \mathbf{c}_\text{a} \):
\begin{equation}
    \mathbf{z}_{\mathrm{full}} = \mathbf{z}_i + \operatorname{CrossAttn}(\mathbf{z}_i, \mathbf{c}_\text{a}),
\end{equation}
which enables each video token to fully capture the overall audio-driven emotional and contextual cues, thereby improving temporal coherence in portrait animation.
\paragraph{Fine-grained Window Fusion.}
The fine-grained fusion output \( \mathbf{z}_{\mathrm{window}} \) is computed by performing cross-attention within each latent frame \( j \in \{1, \ldots, F\} \), where each spatial video token \( \mathbf{z}_i^j \in \mathbb{R}^{(H \times W) \times D} \) (tokens from the \( j \)-th latent) attends to each corresponding audio token \( \mathbf{c}_\text{a}^j \in \mathbb{R}^{(r_f \times N) \times D} \). The results are then concatenated along the frame dimension:
\begin{equation}
   \mathbf{z}_{\mathrm{window}} = \mathbf{z}_i + \operatorname{Concat}\left( \operatorname{CrossAttn}(\mathbf{z}_i^j, \mathbf{c}_\text{a}^j) \right),
\end{equation}
which accurately models the correspondence between lip shapes and local articulation in audio, thus improving the alignment between speech and lip motion.
\paragraph{Hybrid Fusion Strategy.}
Finally, we combine the two fusion branches via a weighted interpolation:
\begin{equation}
\mathbf{z}_{\mathrm{hybrid}} = \alpha(i) \cdot \mathbf{z}_{\mathrm{window}} + (1 - \alpha(i)) \cdot \mathbf{z}_{\mathrm{full}}
\end{equation}
where the interpolation factor $\alpha(i)$ is defined as:
\begin{equation}
\label{alpha}
\alpha(i) = \frac{w \cdot i}{L} + \delta,
\end{equation}
where $i$ denotes the layer index, $L$ denotes the total number of transformer layers, $w$ denotes a scaling parameter, and $\delta$ denotes a bias term.



The hybrid attention mechanism enhances lip synchronization and temporal alignment, while preserving semantic consistency throughout the entire video.
\subsection{Motion-Frame-Free Training and Inference Strategy for Long Video Generation}
\begin{algorithm}[tb]
\caption{Training Algorithm for RAP}
\label{alg:RAP-training}
{\raggedright \textbf{Require}: Encoded video $\mathcal{E}(\mathbf{V})$, Encoded image \( \mathbf{x}_\text{ref} \), Audio feature \( \mathbf{c}_\text{a} \)
\begin{algorithmic}[1]
\WHILE{not converged}
    \STATE Sample timestep $t \sim \mathrm{Uniform}(\{1, \ldots, T\})$
    \STATE Add noise to $\mathcal{E}(\mathbf{V})$ to get $\mathbf{x}_t'$ at timestep $t$
    \STATE Sample $y \sim \mathrm{Bernoulli}(p = \beta)$
    \STATE $\mathbf{x}_t = y \cdot \mathbf{x}_t'[:, :k, :, :] + (1 - y) \cdot \mathbf{x}_t'[:, -k{:}, :, :]$
    \STATE $\tilde{\mathbf{x}}_t=\text{Concat}(\mathbf{x}_t,\mathbf{x}_\text{ref})$
    \STATE Predict the velocity field $\mathbf{v}_t = \mathcal{M}(\tilde{\mathbf{x}}_t, t, \mathbf{c}_\text{a})$

    \STATE Compute flow matching loss $\mathcal{L}$ according to Eq.~\ref{eq:flow_matching}

    
    \STATE Update $\theta$ by minimizing $\mathcal{L}$ using gradient descent
\ENDWHILE

\end{algorithmic}
}
\end{algorithm}

\begin{algorithm}[tb]
\caption{Inference Algorithm with Flow Matching}
\label{alg:RAP-inference}
{\raggedright \textbf{Require}: Encoded image \( \mathbf{x}_\text{ref} \), Audio feature \( \mathbf{c}_\text{a} \), Generative clip number $N$, VAE decoder $\mathcal{D}$\par}
\begin{algorithmic}[1]
\FOR{$i = 1$ to $N$}
\STATE Initialize Gaussian noisy latent: $\mathbf{\hat{x}}_{i,\mathrm{T}} \sim \mathcal{N}(0, \mathbf{I_d})$
\FOR{$t = \mathrm{T}$ to $1$}
\IF{$i \neq 1$}
\STATE $\mathbf{\hat{x}}_{i,t} = \mathrm{Concat}(\mathbf{\hat{x}}_{i-1,t}[:,-n:], \mathbf{\hat{x}}_{i,t}[:,n:])$
  \ENDIF
  \STATE $\mathbf{\tilde{x}}_{i,t} = \mathrm{Concat}(\mathbf{\hat{x}}_{i,t}, \mathbf{x}_\text{ref})$
  \STATE $\mathbf{\hat{x}}_{i,t-1} = \mathbf{\hat{x}}_{i,t} - \Delta t \cdot \mathcal{M}(\mathbf{\tilde{x}}_{i,t}, t,\mathbf{c}_\text{a} )$
  \ENDFOR
\IF{$i = 1$}
    \STATE $\hat{\mathbf{V}}_i = \mathcal{D}(\mathbf{\hat{x}}_{i,0})$
    
\ELSE
    \STATE $\hat{\mathbf{V}}_i = \mathcal{D}(\mathbf{\hat{x}}_{i,0})[:,r_f \cdot(n-1)+1:]$
\ENDIF
\ENDFOR

\end{algorithmic}
\end{algorithm}

In long-form video generation, prior methods often adopt a motion frame strategy, where the last couple frames of the previous clip are used to guide the next. However, this leads to a distribution mismatch: ground-truth motion frames are used during training, while the model relies on its own generated outputs during inference. As generation proceeds, this mismatch accumulates, degrading temporal consistency and visual quality.

To address this issue, we propose a latent inheritance strategy that uses last $n$ intermediate noisy latents from the previous denoising process to softly guide the next clip.  Unlike hard guidance based on denoised final results, this approach avoids direct error injection and reduces the risk of accumulated artifacts. By propagating context through latent features rather than fixed outputs, our method enables more stable and coherent long-form video generation.

 Nevertheless, applying this strategy to 3D VAE architectures introduces a new challenge. These models typically encode identity from a static initial frame and motion from subsequent dynamic frames. Inherited latents from the previous clip, however, are dynamic and inserted at the start of the next clip—disrupting the original static-dynamic structure and compromising the VAE’s encoding.

To resolve this, we propose a dynamic start training scheme. During training, we randomly sample latent features \( \mathbf{x}_t \) from the original $f$-frame \( \mathbf{x}_t' \), following a probabilistic strategy: with probability $\beta$, from the first $a$ frames (containing both static and dynamic latents), and with probability $1 - \beta$, from the last $k$ frames (purely dynamic latents). As a result, this design encourages the model to handle non-static starting conditions. This adaptation ensures compatibility with inherited latents and improves stability across clips. The detailed training and inference procedure is illustrated in Algorithm~\ref{alg:RAP-training} and Algorithm~\ref{alg:RAP-inference}.
\section{Experiments}
\begin{figure*}[ht]
  \centering
    \includegraphics[width=\linewidth]{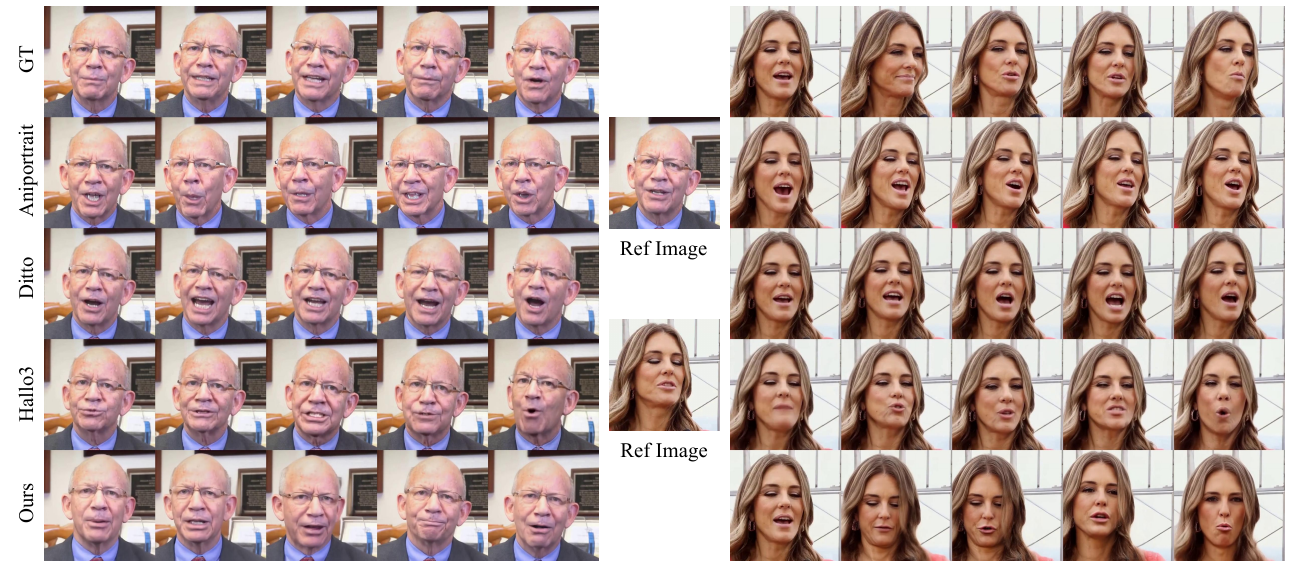}
  \caption{Qualitative comparison with existing approaches on
HDTF and VFHQ dataset. \textbf{Videos are available in the Supplementary Material.}}
  \label{fig:top_double}
\end{figure*}
\begin{figure}[t]
    \includegraphics[width=\linewidth]{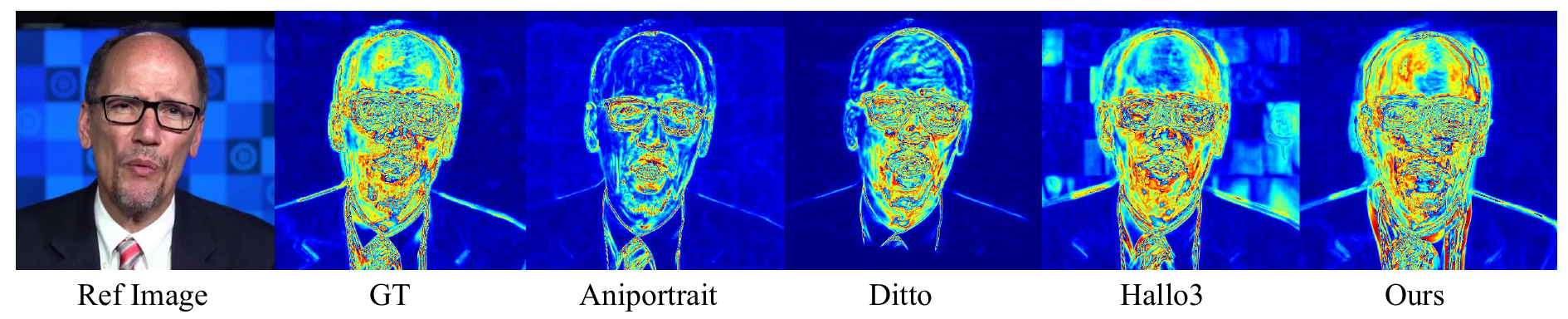}
    \caption{Comparison of temporal consistency and visual drift. Warmer colors denote larger motion amplitude. Our method exhibits minimal background flicker and shift, while preserving significant facial motion. }
    \label{fig:color-compare}
\end{figure}
\subsection{Datasets and Evaluation Metrics}
We constructed the training set from AVSpeech~\cite{ephrat2018looking}, HDTF~\cite{zhang2021flow}, VFHQ~\cite{xie2022vfhq}, and our own collected data. For preprocessing, we first applied face detection to crop each video frame and discarded samples with resolution below $480 \times 480$. All remaining frames were resized to $512 \times 512$. We also used a lip-sync consistency metric to remove samples with poor alignment between lip motion and speech. In parallel, we extracted clean human speech from audio using an audio separation tool. After this pipeline, we obtained 222.6 hours of high-quality paired video and audio data.
For evaluation, we sampled 75 videos per dataset from HDTF and VFHQ. We compared our method with recent state-of-the-art approaches, including SadTalker~\cite{zhang2023sadtalker}, Aniportrait~\cite{wei2024aniportrait}, EchoMimic~\cite{chen2025echomimic}, Ditto~\cite{li2024ditto}, and Hallo3~\cite{cui2025hallo3}.

We employ multiple metrics to comprehensively evaluate our method. For visual quality, we adopt Fréchet Inception Distance (FID)~\cite{heusel2017gans} to measure the distributional discrepancy between generated and real video frames, and Fréchet Video Distance (FVD)~\cite{unterthiner2019fvd} to further capture temporal consistency in video generation. To assess the accuracy and smoothness of audio-visual synchronization, we incorporate Sync-C~\cite{chung2016out}, which reflects how well the lip motion aligns with the speech content, and Sync-D, which captures the temporal stability of lip dynamics throughout the video. Inference efficiency is reported in FPS (frames per second). All evaluations are conducted on a single NVIDIA A800 GPU.

\subsection{Implement Details}
The model was trained on $32$ NVIDIA A800 GPUs using the Adam optimizer.
The input video consists of 121 frames with a spatial resolution of $512 \times 512$. During training, we randomly select either the first 81 frames (static + dynamic) or the last 88 frames (dynamic only) with a 1:1 probability.
Meanwhile, we set a $10\%$ audio dropout to fit the inference classifier-free-guidance (CFG). In training, the batch size of each GPU is $4$ and the learning rate is $1 \times 10^{-5}$. During inference, we set the CFG scale to 5 to preserve the effectiveness of audio-driven control, and use a latent overlap $n=3$ to maintain generation continuity.
Our method requires only 8 GB of GPU memory during inference.

\subsection{Comparison with State-of-the-Art}
\subsubsection{Quantitative Evaluation.}
The quantitative results are reported in Table~\ref{tab:hdtf_results} and Table~\ref{tab:vfhq_results}.
Our method achieves state-of-the-art performance on FVD, Sync-C, and Sync-D, demonstrating strong temporal coherence and superior audio-visual synchronization. While the FID score is slightly inferior to the best-performing baseline—primarily due to the use of highly compressed latent representations that can affect low-level texture fidelity—the gap remains marginal. Additionally, our method runs at real-time inference speed while preserving high perceptual quality.

\begin{table}[ht]
\centering
\resizebox{\linewidth}{!}{%
\begin{tabular}{lccccc}
\toprule
\textbf{Method} &  \textbf{FID$\downarrow$} & \textbf{FVD$\downarrow$} & \textbf{Sync-C$\uparrow$} & \textbf{Sync-D$\downarrow$} & \textbf{FPS$\uparrow$} \\
\midrule
SadTalker    & 21.58 & 207.67 & \underline{4.60} & \underline{9.21}  & 2.17 \\
Aniportrait  & 19.83 & 242.29 & 1.89 & 11.91 & 0.69 \\
EchoMimic    & \textbf{9.00}  & \underline{155.71} & 3.56 & 10.22 & 0.81 \\
Ditto         & 12.35 & 199.13 & 3.57 & 10.49  & \textbf{45.04} \\
Hallo3       & 15.95 & 160.94 & 3.18 & 10.72 & 0.16 \\
\midrule
Ours               & \underline{10.24} & \textbf{122.95} & \textbf{4.85} & \textbf{8.85}  & \underline{42.41} \\
\bottomrule
\end{tabular}
}
\caption{Quantitative comparison on the HDTF dataset.}
\label{tab:hdtf_results}
\end{table}

\begin{table}[ht]
\centering
\resizebox{\linewidth}{!}{%
\begin{tabular}{lccccc}
\toprule
\textbf{Method} &  \textbf{FID$\downarrow$} & \textbf{FVD$\downarrow$} & \textbf{Sync-C$\uparrow$} & \textbf{Sync-D$\downarrow$} & \textbf{FPS$\uparrow$} \\
\midrule
SadTalker    & 29.80 & 191.81 & \underline{4.49} & \underline{8.78}  & 1.60 \\
Aniportrait  & 36.58 & 352.94 & 1.62 & 11.73 & 0.67 \\
EchoMimic    & 24.69 & 193.45 & 2.93 & 10.30 & 0.79 \\
Ditto         & 27.67 & 254.05 & 3.31 & 10.26  & \textbf{41.24} \\
Hallo3       & \underline{23.45} & \underline{171.00} & 4.19 & 9.60 & 0.11 \\
\midrule
Ours               & \textbf{22.68} & \textbf{159.93} & \textbf{4.78} & \textbf{8.40}  & \underline{39.87} \\
\bottomrule
\end{tabular}
}
\caption{Quantitative comparison on the VFHQ dataset.}
\label{tab:vfhq_results}
\end{table}

\subsubsection{Qualitative Evaluation.}

Figure~\ref{fig:top_double} shows visual results generated under identical audio and reference image conditions across different methods.
Our method produces lip movements that are highly consistent with the ground truth, reinforcing its advantage in audio-visual alignment. Furthermore, it generates more diverse facial expressions and exhibits a wider range of motion, resulting in more vivid and expressive portrait animations. In contrast, several baseline methods tend to limit motion amplitude to ensure frame stability and temporal smoothness, which often leads to visually static or less engaging results. By contrast, our method maintains a better balance between temporal consistency and motion expressiveness, enabling high-quality generation that remains responsive to audio dynamics. Meanwhile, Figure~\ref{fig:color-compare} presents the accumulated inter-frame difference map, where warmer areas correspond to larger motion amplitudes. RAP achieves more expressive facial motions with relatively stable background performance. However, other methods either exhibit significant background flicker or produce nearly static characters with only minor local movements.


\begin{figure}[t]
    \centering
    \includegraphics[width=1\linewidth]{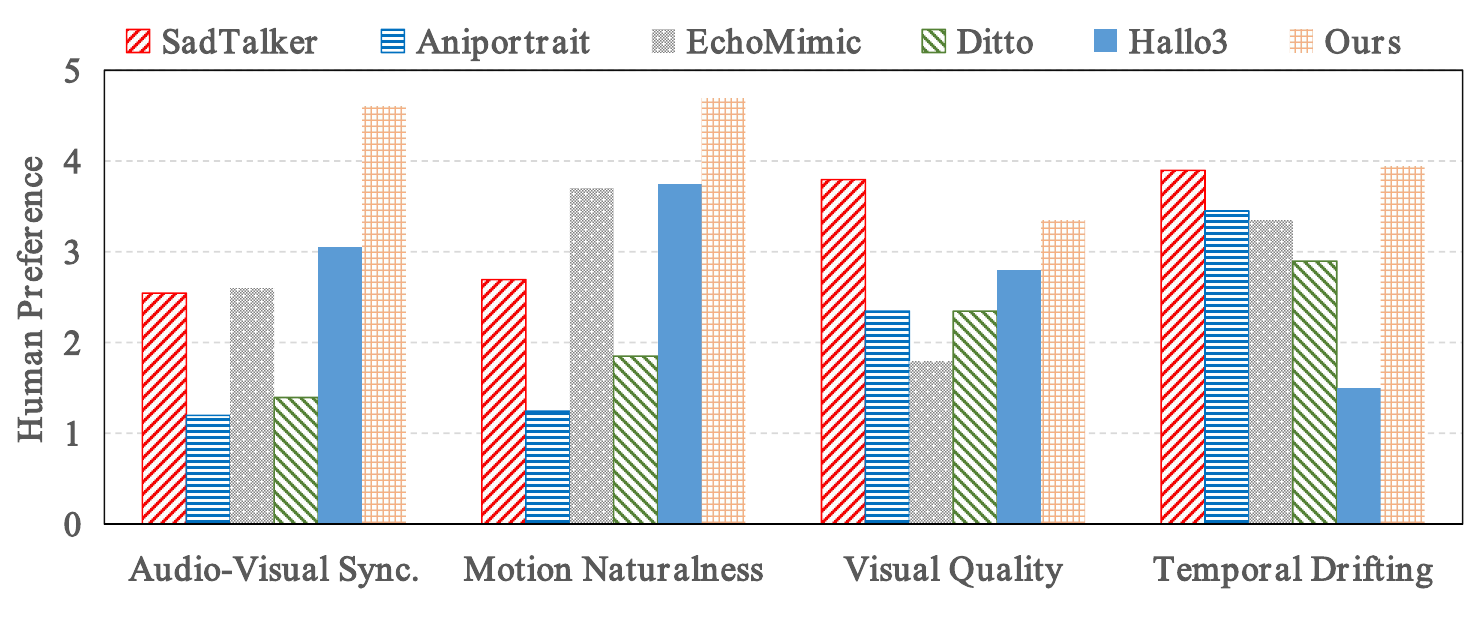} 
    \caption{Human preferences among RAP and baselines.}
    \label{fig:human evaluation}
\end{figure}
\subsubsection{Human Evaluation.}
To assess the perceptual quality of the generated videos, we conducted a human study targeting three key dimensions: audio-visual synchronization, naturalness of human body movement, visual quality and resistance to long-term drifting. We selected 40 video samples with varying durations to reflect different temporal challenges, including 25 short-length clips (10--20 seconds), and 15 extended clips (over 2 minutes). 

The evaluation involved 127 participants (44.1\% aged 18--25, 37.0\% aged 25--35,18.9\% aged 35--50; 40.2\% male, 59.8\% female), of whom 73.2\% had prior experience with AIGC tools or generative video systems. Participants rated each clip on a 5-point Likert scale in terms of audio-visual synchronization, motion naturalness, visual quality, and robustness to temporal drifting. All videos were presented in randomized order to mitigate ordering bias.

As shown in Figure~\ref{fig:human evaluation}, our approach achieved the highest ratings in terms of audio-visual synchronization, motion naturalness, and robustness to temporal drifting.
\subsection{Ablation Studies}
\subsubsection{Hybrid Attention.}


We conduct a comprehensive ablation study comparing four variants: Full-Attention, Window-Attention, a two-stage approach that applies Full-Attention during initial training followed by Window-Attention for fine-tuning, and our proposed Hybrid Attention mechanism. Among these, only the Hybrid Attention effectively integrates both granularities of control simultaneously. It significantly improves lip-audio alignment while maintaining overall motion consistency and coherence. Additionally, it simplifies training by requiring only a single stage. Detailed results are presented in Table~\ref{tab:HA ablation}.


\begin{table}[t]
\centering
\resizebox{\linewidth}{!}{%
\begin{tabular}{lcccc}
\toprule
\textbf{Method} & \textbf{FID$\downarrow$} & \textbf{FVD$\downarrow$} & \textbf{Sync-C$\uparrow$} & \textbf{Sync-D$\downarrow$}  \\
\midrule
Full-Attention    &  14.49 & 196.29 & 2.13 & 11.42 \\
Window-Attention   &  12.35 & 176.16 & \textbf{5.03} & \textbf{8.68}  \\
Full+Window(2stages)      &  \underline{11.89} & \underline{166.02} & 4.81 & 8.86 \\
Hybrid Attention          &  \textbf{10.24} & \textbf{122.95} & \underline{4.85} & \underline{8.85} \\
\bottomrule
\end{tabular}
}
\caption{Comparison of different audio injection methods on HDTF dataset.}
\label{tab:HA ablation}
\end{table}

We further ablate $w$ and $\delta$ in Eq.~\ref{alpha}, as shown in table~\ref{tab:parameter ablation}. We set \( w = 1 \) and \( \delta = 0 \) to achieve the best visual quality while preserving high audio-visual synchronization accuracy.

\begin{table}[t]
\centering
\resizebox{\linewidth}{!}{%
\begin{tabular}{ll|cccc}
\toprule
\multicolumn{2}{c}{\textbf{Settings}} & \textbf{FID$\downarrow$} & \textbf{FVD$\downarrow$} & \textbf{Sync-C$\uparrow$} & \textbf{Sync-D$\downarrow$} \\
\midrule
\multirow{5}{*}{$w=0$}
 & $\delta=0$      &  14.49 & 196.29 & 2.13 & 11.42 \\
 & $\delta=0.25$   &  11.31 & 136.59 & 3.48 & 9.94 \\
 & $\delta=0.5$    &  10.46 & 126.42 & 4.80 & 8.87 \\
 & $\delta=0.75$   &  10.92& 138.97 & \underline{4.91} & \underline{8.78} \\
 & $\delta=1$      & 12.35  & 176.16 & \textbf{5.03} & \textbf{8.68} \\
\midrule
  $w=-1$     &  \multirow{4}{*}{$\delta=0.5-w/2$}   &  10.91 & 128.69 & 4.90 & 8.79 \\
  $w=-0.5$   &      &  10.79 & 127.32 & 4.88 & 8.80 \\
  $w=0.5$ &       &  \underline{10.41} & \underline{124.73} & 4.82 & 8.83 \\
  $w=1$  &        &  \textbf{10.24} & \textbf{122.95} & 4.85 & 8.85 \\

\bottomrule
\end{tabular}
}
\caption{Comparison of the effects of varying $w$ and $\delta$ on hybrid attention.}
\label{tab:parameter ablation}
\end{table}
\subsubsection{Training and Inference Strategy.}
Vanilla motion-frame-based approaches directly inject previously generated content as a condition for the subsequent generation process and are trained in a corresponding manner. This leads to a strong dependency on the input motion frames and causes the model to inherit and accumulate errors from them. In contrast, our proposed training and inference strategy utilizes the preceding generated results solely to guide the denoising process of the next generative clip, thereby circumventing the conventional teacher-forcing dilemma. As shown in Figure~\ref{fig:headimage-V1} (a), with extended inference duration, the results from the motion-frame baseline rapidly develop conspicuous artifacts that progressively accumulate. Conversely, RAP's results exhibit no significant degradation as the duration increases. Our tests demonstrate that a one-hour-long video result can maintain the same quality as its initial segments.
Furthermore, our mixed static-dynamic training is specifically designed to complement the aforementioned inference scheme. Unlike traditional methods that train exclusively on static-to-dynamic pairs, our strategy also incorporates dynamic-to-dynamic pairs. This approach resolves the inheritance inconsistency issue that typically emerges from the second window onward, thus achieving superior transitional performance, as Figure~\ref{fig:headimage-V1} (b) illustrated.
\begin{figure}[t]
    \includegraphics[width=\linewidth]{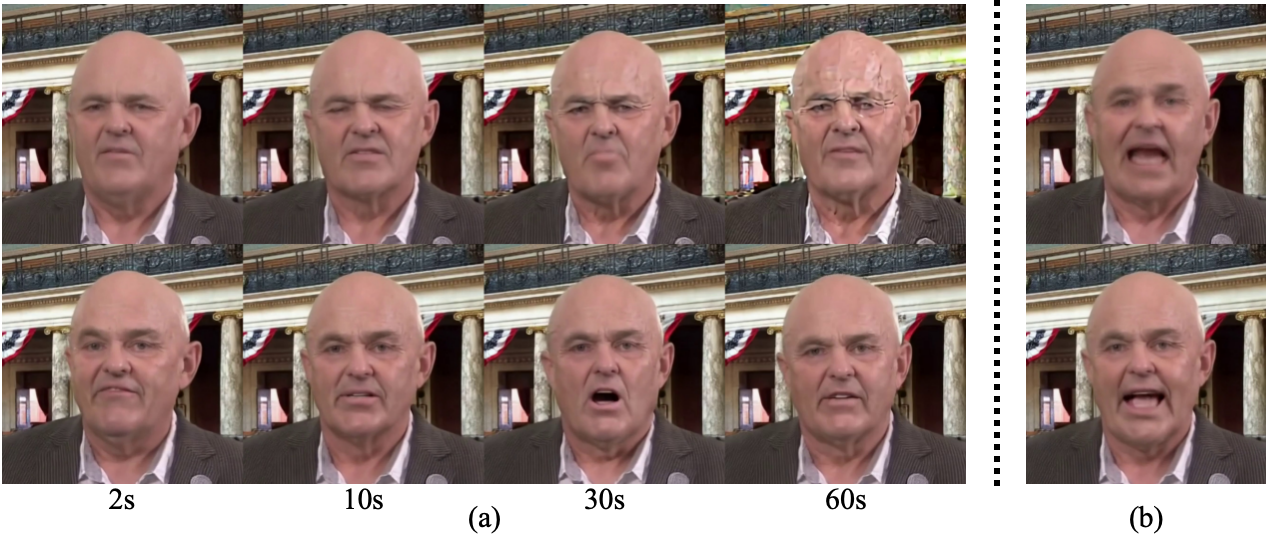}
\caption{Comparison between inference strategies: subfigure (a) top shows the motion-frame-guided inference, while subfigure (a) bottom shows the inference strategy adopted by RAP. Comparison between training strategies: subfigure (b) top shows training from static latent only, while subfigure (b) bottom shows our hybrid training strategy.
}
    \label{fig:headimage-V1}
\end{figure}
\section{Conclusion}
In this work, we propose RAP, a real-time audio-driven portrait animation framework. By introducing a hybrid attention mechanism and a static-dynamic joint training-inference strategy, RAP achieves precise alignment between audio and visual content under highly compressed representations, enabling the real-time generation of natural and coherent long-term portrait animations.

Under rapid motion scenarios, the use of a high compression ratio VAE may still lead to motion blur and ghosting artifacts due to latent information loss, limiting the fidelity of generated results. Furthermore, extending the framework to real-time multi-speaker conversations and dynamic scene generation remains an important direction for future work. In addition, exploring the applicability of our training and inference strategy to other modality-guided portrait animation tasks, and more broadly to general video generation, is a promising avenue for future research.

\bibliography{aaai2026}
\clearpage
\appendix
\section{Appendix}
\subsection{Dataset processing}
The training data used in this work includes AVSpeech~\cite{ephrat2018looking}, HDTF~\cite{zhang2021flow}, VFHQ~\cite{xie2022vfhq}, and our own collected video dataset. Since AVSpeech and our collected data contain a large number of low-quality samples (e.g., low resolution or audio-visual misalignment), we apply a series of preprocessing steps to ensure the quality of audio-driven face generation. Specifically, we first compute the Sync-C and Sync-D metrics for each video and retain only those with Sync-C $>$ 1 and Sync-D $<$ 13. We then perform face detection on the first frame and discard videos in which the detected face region is smaller than $480 \times 480$. For the remaining videos, the face bounding box is expanded by a factor of two to ensure that the face occupies approximately half of the cropped frame width, followed by resizing to $512 \times 512$ to meet the network input requirements. All audio tracks are converted to mono and undergo offline voice extraction, with the resulting clean speech saved as \texttt{.pt} files for efficient training. Moreover, we discard videos shorter than 5 seconds to ensure sufficient sequence length for training with both static and dynamic frames. After preprocessing, the total durations of the datasets are as follows: AVSpeech 121.6 hours, our own dataset 81.39 hours, HDTF 6.73 hours, and VFHQ 12.86 hours.

\begin{figure}[h]
    \centering
    \includegraphics[width=\linewidth]{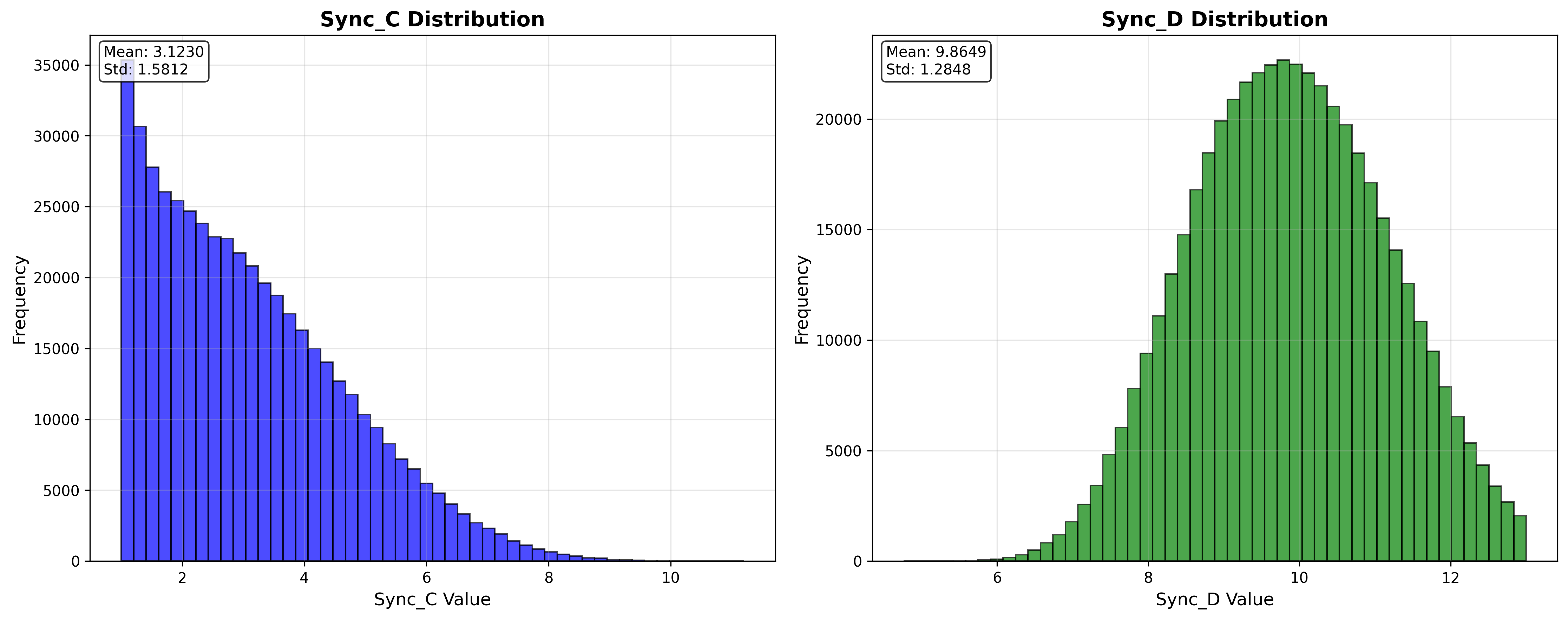}
    \caption{Distribution of Sync-C and Sync-D in raw datasets.}
     
    \label{fig:distribution}
\end{figure}
\subsection{Experiments}
\subsubsection{Experimental Setup Details.}
In RAP, we adopt the LTXVAE~\cite{hacohen2024ltx} together with the Wan2.1~\cite{wan2025wan} T2V model containing 1.3B parameters. Specifically, the raw video frames $\mathbf{V} \in \mathbb{R}^{3 \times 121 \times 512 \times 512}$ are compressed by LTX-VAE with a spatial-temporal compression factor of $(8, 32, 32)$ to get $\mathbf{x}_0 \in \mathbb{R}^{128 \times 16 \times 16 \times 16}$. We add Gaussian noise to $\mathbf{x}_0$ based on a sampled timestep $t \sim \mathcal{U}(0, T)$, obtaining the noisy latent $\mathbf{x}_t'\in \mathbb{R}^{128 \times 16 \times 16 \times 16}$, from which we randomly sample latent features $\mathbf{x}_t \in \mathbb{R}^{128 \times 11 \times 16 \times 16}$  following a probabilistic strategy: with probability 0.5, from the first 11 frames (containing both static and dynamic latents), and with probability 0.5, from the last 11 frames (purely dynamic latents). To preserve identity, we extract the corresponding latent features $\mathbf{x}_{\text{ref}}$ from the reference image and concatenate them with $\mathbf{x}_t$ along the channel dimension, yielding the fused input $\tilde{\mathbf{x}}_t \in \mathbb{R}^{256 \times 11 \times 16 \times 16}$.

This fused latent $\tilde{\mathbf{x}}_t$ is reshaped and processed by a patchify layer with kernel size $(1, 1, 1)$, resulting in a token sequence $\mathbf{z}_0 \in \mathbb{R}^{(11 \times 16 \times 16) \times 1536}$ to match the input dimension of the Transformer. We adopt a 30-layer Transformer with 12 attention heads and a feed-forward network (FFN) hidden dimension of 8960 to enhance expressive capacity.

The output sequence is then depatchified to reconstruct the latent $\hat{\mathbf{x}}_0 \in \mathbb{R}^{128 \times 11 \times 16 \times 16}$, and decoded by the VAE decoder to generate video frames $\hat{\mathbf{V}} \in \mathbb{R}^{3 \times 81 \times 512 \times 512}$.
\subsubsection{The impact of CFG scale.}

Classifier-free-guidance (CFG) is a crucial technique for enhancing controllability during model inference. We conduct an ablation study on different CFG scales, and the results are shown in the table~\ref{tab:cfg} below.
\begin{table}[ht]
\centering
\resizebox{\linewidth}{!}{%
\begin{tabular}{lccccc}
\toprule
\textbf{CFG scale} &  \textbf{FID$\downarrow$} & \textbf{FVD$\downarrow$} & \textbf{Sync-C$\uparrow$} & \textbf{Sync-D$\downarrow$}  \\
\midrule
2    & 11.50 & 123.94 & 4.61 & 9.07   \\
4  & 10.33 & \textbf{120.48} & \underline{4.83} & \textbf{8.85}  \\
5    & \underline{10.22} & \underline{122.95} & \textbf{4.85} & \textbf{8.85}  \\
6         & \textbf{10.08} & 126.55 & 4.79 & \underline{8.91}  \\
8       & 10.89 & 130.31 & 4.71& 8.98  \\
\bottomrule
\end{tabular}
}
\caption{Ablation of CFG scale on the HDTF dataset. We set it to 5 to balance visual quality and audio-visual synchronization performance.}
\label{tab:cfg}
\end{table}

\begin{table}[ht]
\centering
\resizebox{\linewidth}{!}{%
\begin{tabular}{lccccc}
\toprule
\textbf{overlap length} & \textbf{FID$\downarrow$} & \textbf{FVD$\downarrow$} & \textbf{Sync-C$\uparrow$} & \textbf{Sync-D$\downarrow$} & \textbf{FPS$\uparrow$} \\
\midrule
$n=1$           & 10.61 & 125.63 & \textbf{4.89} & \textbf{8.82} & \textbf{52.96} \\
$n=2$           & 10.38 & \underline{123.58} & \underline{4.86} & 8.87 & \underline{46.71} \\
$n=3$ (Ours)    & \underline{10.24} & \textbf{122.95} & 4.85 & \underline{8.85} & 42.41 \\
$n=4$           & \textbf{10.22} & 123.67 & 4.83 & 8.89 & 38.12 \\
\bottomrule
\end{tabular}
}
\caption{Ablation of overlap length on the HDTF dataset.}
\label{tab:ablation_n}
\end{table}
\subsubsection{The impact of overlap length.}
In our approach, we adopt a temporal guidance strategy, where the last $n$ latent frames

\noindent of the preceding clip are reused as the first $n$ latents of the next clip to guide generation. We conduct an ablation study on the number of overlapping latent frames $n$ to investigate its impact on generation quality, audio-visual synchronization, and inference speed. In the final setting, we choose $n=3$ as it achieves the best trade-off between visual fidelity, sync accuracy, and efficiency.

\subsubsection{More qualitative results.}
\begin{figure*}[th]
    \centering
    \subfloat[]{%
        \includegraphics[width=0.5\linewidth]{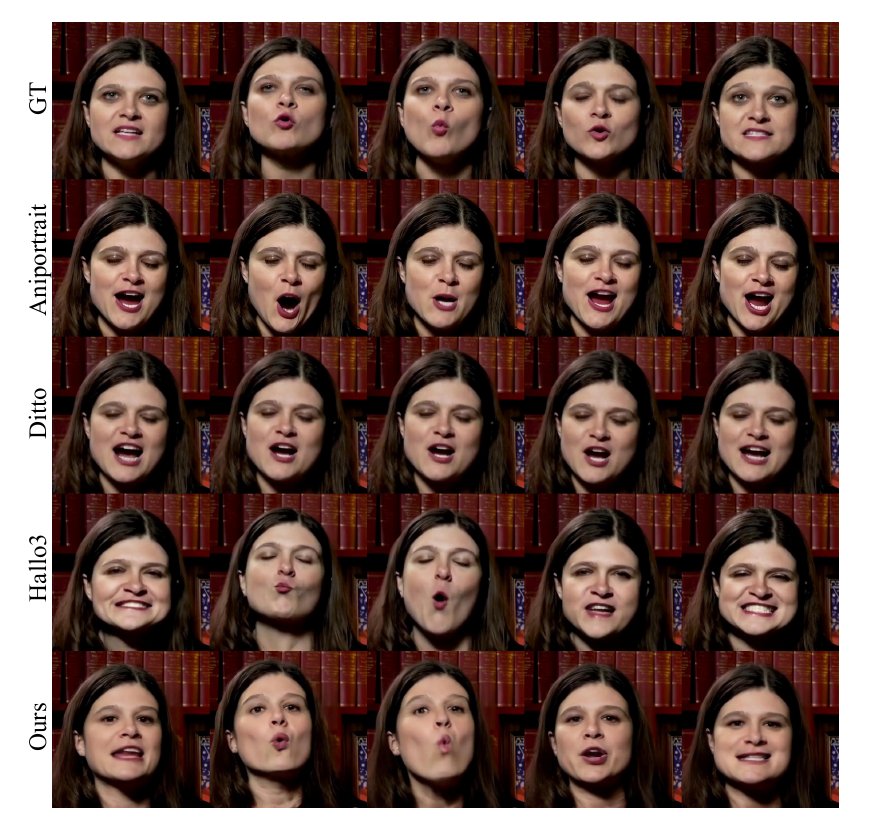}   
        \label{}
        }%
    \subfloat[]{
        \includegraphics[width=0.5\linewidth]{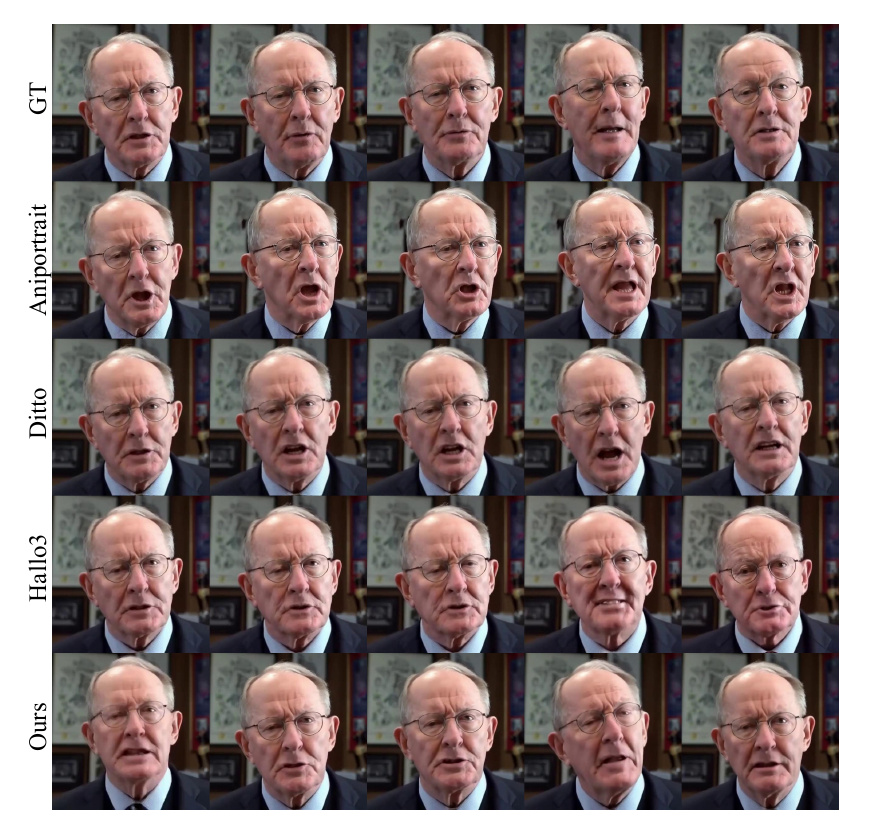}
        \label{}
        }
    \caption{Qualitative results on HDTF dataset}
    \label{fig:1}
\end{figure*}

\begin{figure*}[th]
    \centering
    \subfloat[]{%
        \includegraphics[width=0.5\linewidth]{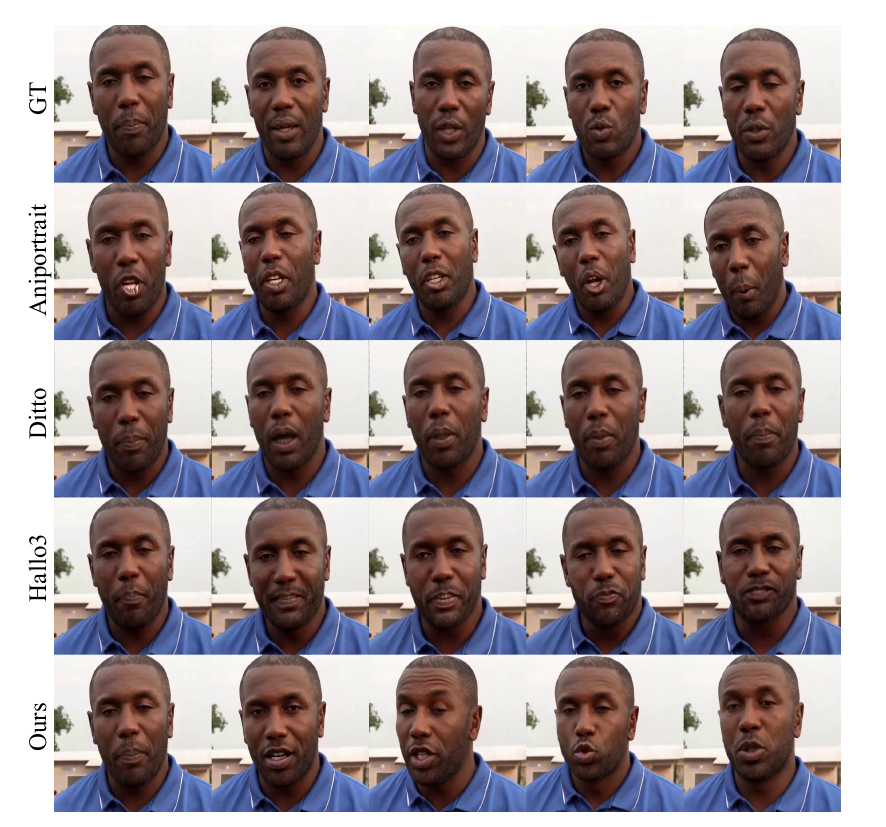}   
        \label{}
        }%
    \subfloat[]{
        \includegraphics[width=0.5\linewidth]{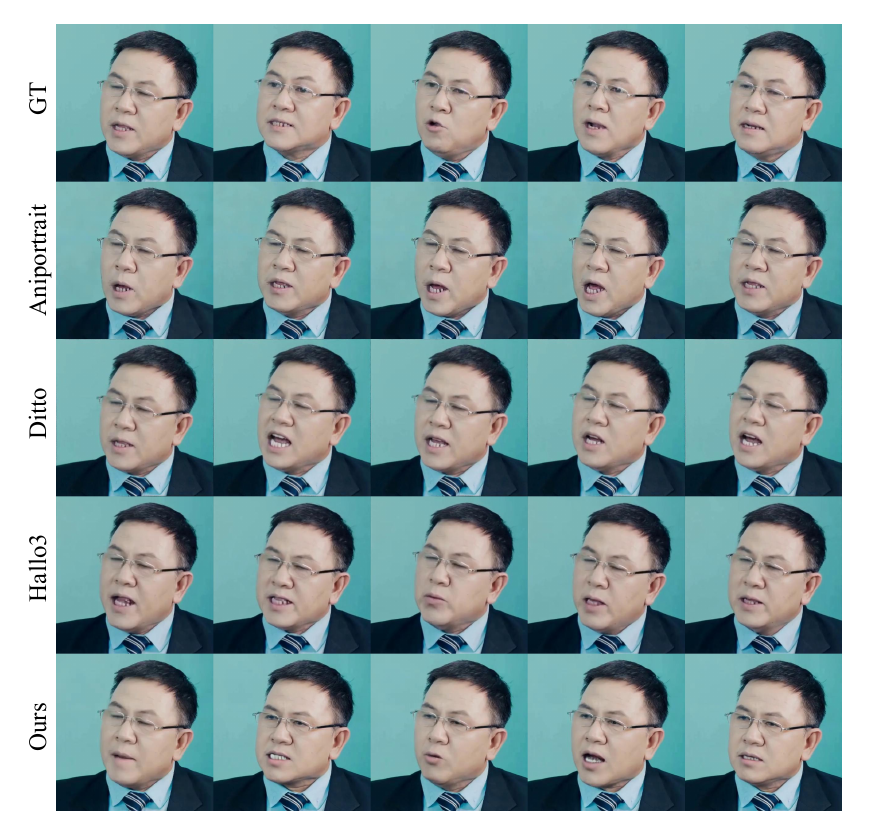}
        \label{}
        }
    \caption{Qualitative results on VFHQ dataset}
    \label{fig:2}
\end{figure*}

More qualitative results are illustrated in Figures~\ref{fig:1},~\ref{fig:2}

\end{document}